\documentclass[]{article}
\usepackage{authblk}

\usepackage{float}
\usepackage{graphicx}
\usepackage{amsmath}
\usepackage{adjustbox}
\usepackage{threeparttable}
\usepackage{hyperref}
\usepackage{multirow}

\begin{document}

\title{The COVID-19 Social Media Infodemic} 

\author[1]{Matteo Cinelli}
\author[2,1,3]{Walter Quattrociocchi\thanks{corresponding author: w.quattrociocchi@unive.it}}
\author[4]{Alessandro Galeazzi}
\author[5]{Carlo Michele Valensise} 
\author[1]{Emanuele Brugnoli}
\author[2]{Ana Lucia Schmidt}  
\author[6]{Paola Zola} 
\author[2,1]{Fabiana Zollo}  
\author[1,3]{Antonio Scala}

\affil[1]{CNR-ISC, Roma}
\affil[2]{Università Ca’ Foscari di Venezia}
\affil[3]{Big Data in Health Society, Roma}
\affil[4]{Università di Brescia}
\affil[5]{Politecnico di Milano}
\affil[6]{CNR-IIT, Pisa}

\date{}

\maketitle

\begin{abstract}
We address the diffusion of information about the COVID-19 with a massive data analysis on Twitter, Instagram, YouTube, Reddit and Gab. We analyze engagement and interest in the COVID-19 topic and provide a differential assessment on the evolution of the discourse on a global scale for each platform and their users. We fit information spreading with epidemic models characterizing the basic reproduction numbers $R_0$ for each social media platform. Moreover, we characterize information spreading from questionable sources, finding different volumes of misinformation in each platform. However, information from both reliable and questionable sources do not present different spreading patterns. Finally, we provide platform-dependent numerical estimates of rumors' amplification.
\end{abstract}

%{\bf  We model information diffusion regarding the COVID-19 and estimate rumors' amplification in various social media platforms.}

\section*{Introduction}
The World Health Organization (WHO) defined the SARS-CoV-2 virus (initially known as 2019-nCoV) outbreak as a severe global threat
\footnote{\href{https:www.who.int/emergencies/diseases/novel-coronavirus-2019/technical-guidance/naming-the-coronavirus-disease-(covid-2019)-and-the-virus-that-causes-it}{WHO Link: Naming the coronavirus disease (COVID-19) and the virus that causes it.}}. As foreseen already in 2017 by the global risk report of the World Economic forum, global risks are interconnected; in particular, the case of the COVID-19 epidemic (the infectious disease caused by the most recently discovered human coronavirus) is showing the critical role of information diffusion in a disintermediated news cycle \cite{quattrociocchi2017part}.

The term \textit{infodemic} \cite{whositrep13,zarocostas2020fight} has been coined to outline the perils of misinformation phenomena during the management of virus outbreaks\footnote{\href{https://www.who.int/dg/speeches/detail/director-general-s-remarks-at-the-media-briefing-on-2019-novel-coronavirus---8-february-2020}{WHO Link: Director-General’s remarks at the media briefing on 2019 novel coronavirus on 8 February 2020}} \cite{mendoza2010twitter, starbird2014rumors}, since it could even speed up the epidemic process by influencing and fragmenting social response \cite{Kim2019}.  
As an example, CNN has recently anticipated a rumor about the 
possible lock-down of Lombardy (a region in northern Italy) to prevent pandemics\footnote{\href{https://edition.cnn.com/2020/03/08/europe/italy-coronavirus-lockdown-europe-intl/index.html}{CNN Link: Italy prohibits travel and cancels all public events in its northern region to contain coronavirus}}, publishing the news hours before the official communication from the Italian Prime Minister. As a result, people overcrowded trains and airports to escape from Lombardy toward the southern regions before the lock-down was in place, disrupting the government initiative aimed to contain the epidemics and potentially increasing contagion.
Thus, an important research challenge is to determine how people seek or avoid information and how those decisions affect their behavior \cite{sharot2020people}, particularly when the news cycle -- dominated by the disintermediated diffusion of information -- alters the way information is consumed and reported on. 
The case of the COVID-19 epidemic shows the critical impact of this new information environment. 

The information spreading can strongly influence people behavior and alter the effectiveness of the countermeasures deployed by governments.
To this respect, models to forecast virus spreading are starting to account for the behavioral response of the population with respect to public health interventions and the communication dynamics behind content consumption \cite{shaman2013real,Kim2019,viboud2019future}.

Social media platforms such as Youtube and Twitter provide direct access to an unprecedented amount of content and may amplify rumors and questionable information. Taking into account users' preferences and attitudes, algorithms mediate and facilitate content promotion and thus information spreading \cite{kulshrestha2017quantifying}. This shift of paradigm profoundly impacts the construction of social perceptions \cite{schmidt2017anatomy} and the framing of narratives; it influences policy-making, political communication, as well as the evolution of public debate \cite{starnini2016emergence,schmidt2018polarization} especially when issues are controversial \cite{del2016spreading}.
Indeed, users online tend to acquire information adhering to their worldviews \cite{bessi2015science}, to ignore dissenting information \cite{zollo2017debunking,baronchelli2018emergence} and to form polarized groups around shared narratives \cite{del2016echo,bail2018exposure}. Furthermore, when polarization is high, misinformation might easily proliferate \cite{vicario2019polarization,wardle2017information}. Some studies pointed out that fake news and inaccurate information may spread faster and wider than fact-based news \cite{vosoughi2018spread}. However, this effect might be platform-specific.
The definition of ``Fake News" may indeed be inadequate since political debate often resorts to label opposite news as unreliable or fake \cite{ruths2019misinformation}.

In this work we provide an in-depth analysis of social dynamics in a time window where narratives and moods in social media related to the COVID-19 have emerged and spread. While most of the studies on misinformation diffusion focus on a single platform \cite{vosoughi2018spread, bovet2019influence, del2016spreading}, the dynamics behind information consumption might be particular to the environment in which they spread on. 
Consequently, in this paper we perform a comparative analysis on five social media platforms (Twitter, Instagram, YouTube, Reddit and Gab) during the COVID-19 outbreak. The dataset includes more than 8 million comments and posts over a time span of 45 days. We analyze user engagement and interest about the COVID-19 topic, providing an assessment of the discourse evolution over time on a global scale for each platform. Furthermore, we model the spread of information with epidemic models, characterizing for each platform its basic reproduction numbers ($R_0$), i.e. the average number of secondary cases (users that start posting about COVID-19) an ``infectious" individual (an individual already posting on COVID-19) will create. In epidemiology, $R_0$ is a threshold parameter, where for $R_0 < 1$ the disease will die out in a finite period of time, while the disease will spread for $R_0>1$. In social media, $R_0>1$ will indicate the possibility of an infodemic.

Finally, coherently with the classification provided by the fact-checking organization Media Bias/Fact Check\footnote{\url{https://mediabiasfactcheck.com} classifies news sources that are considered reliable and news sources that are considered unreliable}, we characterize the spreading of news regarding COVID-19 from questionable sources for all channels but Instagram, finding that mainstream platforms are less susceptible to misinformation diffusion. However, information marked either as reliable or questionable do not present significant differences in the way they spread. 

Our findings suggest that the interaction patterns of each social media platform combined with the peculiarity of the audience of the specific platform play a pivotal role in information and misinformation spreading. We conclude the paper by measuring rumor’s amplification parameters for COVID-19 on each social media platform.

\section{Results and Discussion}
We analyze mainstream platforms such as Twitter, Instagram and YouTube as well as less regulated social media platforms such as Reddit and Gab. Gab is a crowdfunded social media whose structure and features are Twitter-inspired. It performs very little control on content posted; in the political spectrum, its user base is considered to be far-right. Reddit is an American social news aggregation, web content rating, and discussion website based on collective filtering of information.

We perform a comparative analysis of information spreading dynamics around the same argument in different environments having different interaction settings and audiences.
We collect all pieces of content related to COVID-19 from the 1st of January to the 14th of February. 
Data have been collected filtering contents accordingly to a selected sample of Google Trends' COVID-19 related queries such as: \textit{coronavirus}, \textit{coronavirusoutbreak}, \textit{imnotavirus}, \textit{ncov}, \textit{ncov}-19, \textit{pandemic}, \textit{wuhan}. The deriving dataset is then composed of 1,342,103 posts and 7,465,721 comments produced by 3,734,815 users. For more details regarding the data collection refer to SI.

\subsection{Interaction patterns}

\begin{figure}[H]
    \centering
    \includegraphics[width=.95\textwidth]{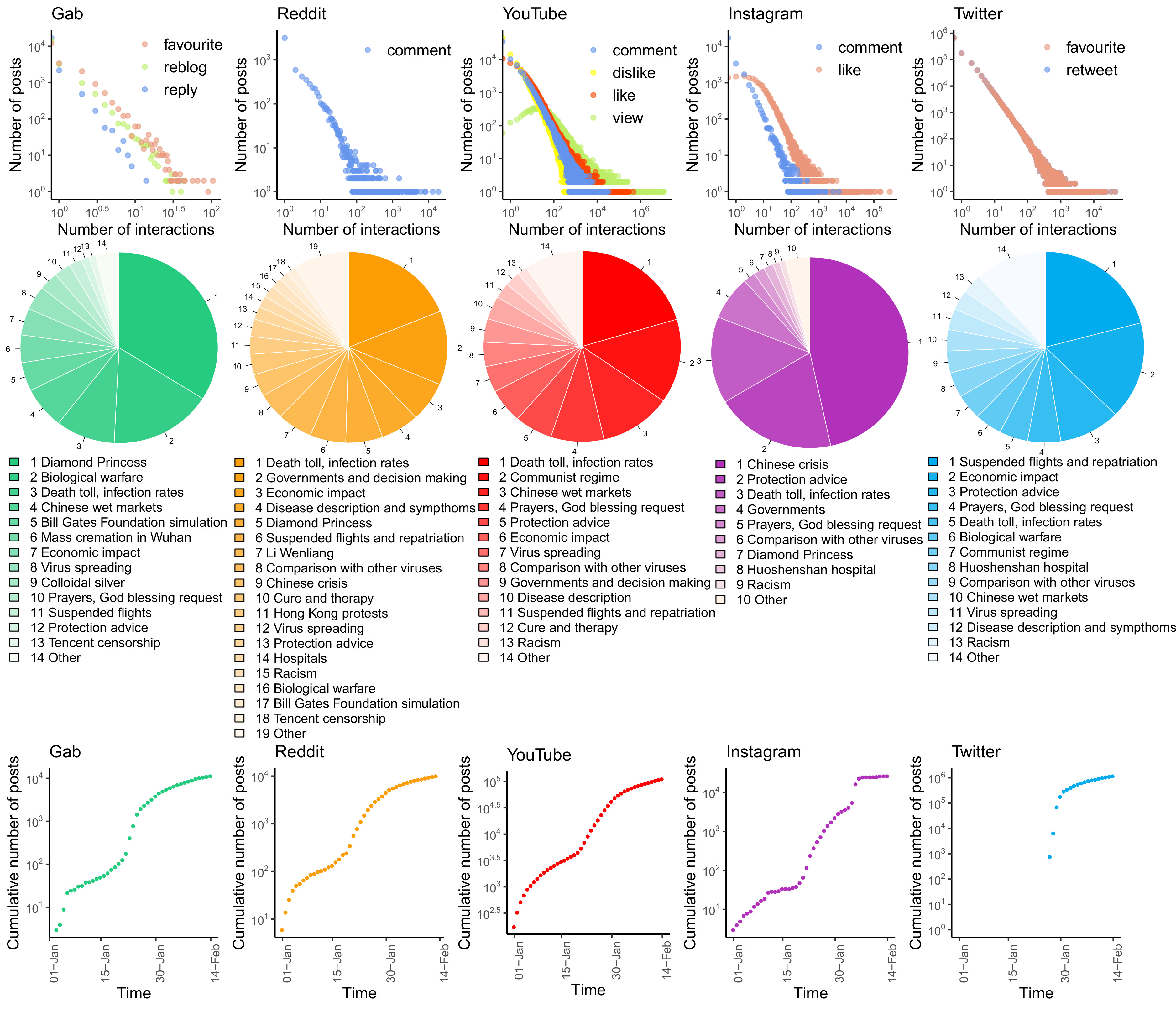}
    \caption{Upper panel: activity (likes, comments, reposts, etc..) distribution for each social media. 
    Middle panel: most discussed topics about COVID-19 on each social media.
    Lower panel: cumulative number of contents produced from the 1$^{st}$ of January to the 14$^th$ of February. Due to limitations in gathering past data using the standard API, the first data point for Twitter is dated January 27$^{th}$.}
    \label{fig:1}
\end{figure}

First, we analyze the interactions (i.e., the engagement) that users have with COVID-19 topics on each platform.
The upper panel of Figure \ref{fig:1} shows users' engagement around the COVID-19. 
Despite the differences among the single platforms, we observe that they all display a rather similar distribution of the users' activity characterized by a long tail. 
This entails that users behave similarly for what concern the dynamics of reactions and content consumption. Indeed, users' interactions with the COVID-19 content present attention patterns similar to any other topic \cite{romero2011differences}. The highest volume of interactions in terms of posting and commenting can be observed on mainstream platforms such as YouTube and Twitter. 
Then, to provide an overview of the debate concerning the virus outbreak, we extract and analyze all topics related to the COVID-19 content by means of Natural Language Processing techniques.
We build word embedding for the text corpus of each platform, i.e. a word vector representation in which words sharing common lexical contexts are in close proximity. Moreover, by running clustering procedures on these vector representations, we separate groups of words and topics that are perceived as more relevant for the COVID-19 debate. For further details see SI.
The results (Figure \ref{fig:1}, middle panel) show that topics are quite similar across each social media platform. Debates range from comparisons to other viruses, requests for God blessing, up to racism, while the largest volume of interaction is related to the lock-down of flights.

Finally, to characterize users engagement with the COVID-19 on the five platforms, we compute the cumulative number of new posts each day (Figure \ref{fig:1}, middle panel). For all platforms, we find a change of behavior around the 20$^{th}$ of January, that is the day that the World Health Organization (WHO) issued its first situation report on the COVID-19\footnote{\href{https://www.who.int/docs/default-source/coronaviruse/situation-reports/20200121-sitrep-1-2019-ncov.pdf?sfvrsn=20a99c10_4}{WHO Link: Novel Coronavirus (2019-nCoV) SITUATION REPORT - 1}}.
The largest increase in the number of posts is the on the 21$^{st}$ of January for Gab, the 24$^{th}$ January for Reddit, the 30$^{th}$ January for Twitter, the 31$^{th}$ January for YouTube and the 5$^{th}$ of February for Instagram. Thus, social media platforms seem to have specific timings for content consumption; such patterns may depend upon the difference in terms of audience and interaction mechanisms (both social and algorithmic) among platforms.

\subsection{Information Spreading}

Efforts to simulate the spreading of information on social media by reproducing real data have mostly applied variants of standard epidemic models \cite{pellis2015eight,Liu2016,Skaza2017,davis2020phase}. Coherently, we analyze the observed monotonic increasing trend in the way new users interact with information related to the COVID-19 by means of epidemic models. Unlike previous works, we do not only focus on models that imply specific growth mechanisms, but also on phenomenological models that emphasize the reproducibility of empirical data \cite{Chowell2017}.  

Most of the epidemiological models focus on the basic reproduction number $R_0$, representing the expected number of infections directly generated by an infected individual for a given time period \cite{Ma2020}. An epidemic is considered to be dangerous if $R_0>1$, -- i.e., if an exponential growth in the number of infections is expected at least in the initial phase. In our case, we try to model the growth in number of people publishing a post on a subject as an infective process, where people can start publishing after being exposed to the topic. While in real epidemics $R_0>1$ highlights the possibility of a pandemic, in our approach $R_0>1$ indicates the possibility of an infodemic. We model the dynamics both with the phenomenological model of \cite{Fisman2013} (from now on referred to as the EXP model) and with the standard SIR (Susceptible, Infected, Recovered) compartmental model \cite{Bailey1975book}. Further details on the modeling approach can be found in Section \ref{subsec:MM-epidemics}.

\begin{figure}[H]
    \centering
    \includegraphics[width=1.0\textwidth]{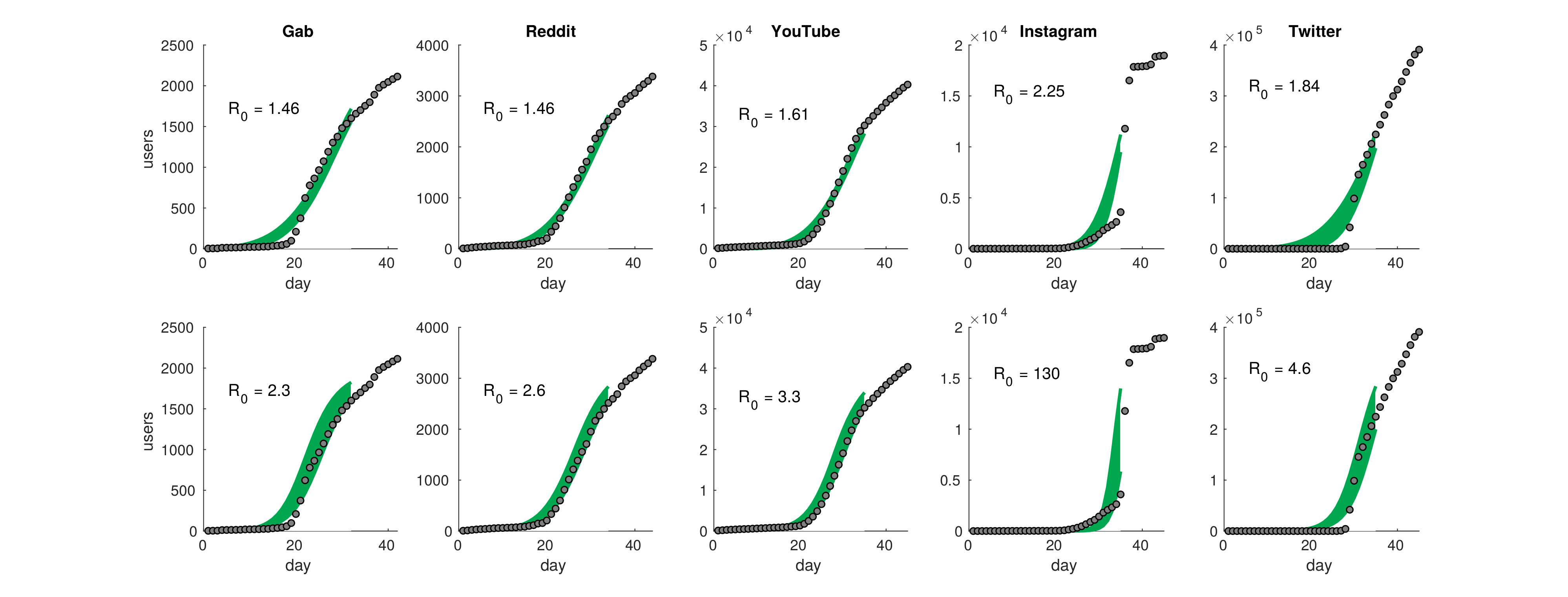}
    \caption{Growth of the number of authors vs time. Time is expressed in number of days since $1^{st}$ Jan $2020$ (day 1). Shaded areas represents $[5\% , 95\%]$ estimates of the models obtained via bootstrapping  least square estimates of the EXP model (upper panels, eq. \protect{\ref{eq:EXP}}) and of the SIR model  (lower panels, eq. \protect{\ref{eq:SIR}}).}
    \label{fig:Epidemic}
\end{figure}

As shown in Figure~\ref{fig:Epidemic}, %and Table \ref{tab:R0} 
each platform has its own basic reproduction number $R_0$. As expected, all the values of $R_0$ are supercritical - even considering confidence intervals (table \ref{tab:DeltaR0}) - signaling the possibility of an infodemic. This observation may facilitate the prediction task of information spreading during critical events. 

While $R_0$ is a good proxy for the engagement rate and a good predictor for epidemic-like information spreading, social contagion phenomena might be in general more complex \cite{centola2010spread,del2017modeling,baumann2020modeling}. For instance, in the case of Instagram, we observe an abrupt jump in the number of new users that cannot be explained with continuous models like the standard epidemic ones; accordingly, the SIR model estimates a value of $R_0\sim 10^2$ that is way beyond what has been observed in any real-world epidemic.

\begin{table}[H]
    \centering
    \begin{tabular}{l|l|l|l|l|l}
        \hline 
        & \bf Gab & \bf Reddit & \bf YouTube & \bf Instagram & \bf Twitter \\ 
        $\mathbf{R_0^{EXP}}$ & $[1.42 , 1.52]$ & $[1.44 , 1.51]$ & $[1.56 , 1.70]$ & $[2.02 , 2.64]$ & $[1.65 , 2.06]$ \\
        $\mathbf{R_0^{SIR}}$ & $[2.2 , 2.5]$ & $[2.4 , 2.8]$ & $[3.2 , 3.5]$ & $[1.1x10^2 , 1.6x10^2]$ & $[4.0 , 5.1]$ \\
        \hline
    \end{tabular}
    \caption{$[5\%,95\%]$ interval of confidence $R_0$ as estimated from bootstrapping the least square fits parameter of the EXP and of the SIR model. Notice that, due to the steepness of the growth of the number of new authors in Instagram, $R_0$ assumes unrealistic values $\sim 10^2$ for the SIR model.}
    \label{tab:DeltaR0}
\end{table}

\subsection{Questionable VS Accurate Information}

We conclude our analysis by comparing the diffusion of questionable and reliable news on each platform. We tag links as reliable or questionable according to the data reported by the independent fact-checking organization Media Bias/Fact Check\footnote{\url{https://mediabiasfactcheck.com}}.

\begin{figure}[H]
    \centering
    \includegraphics[width=.95\textwidth]{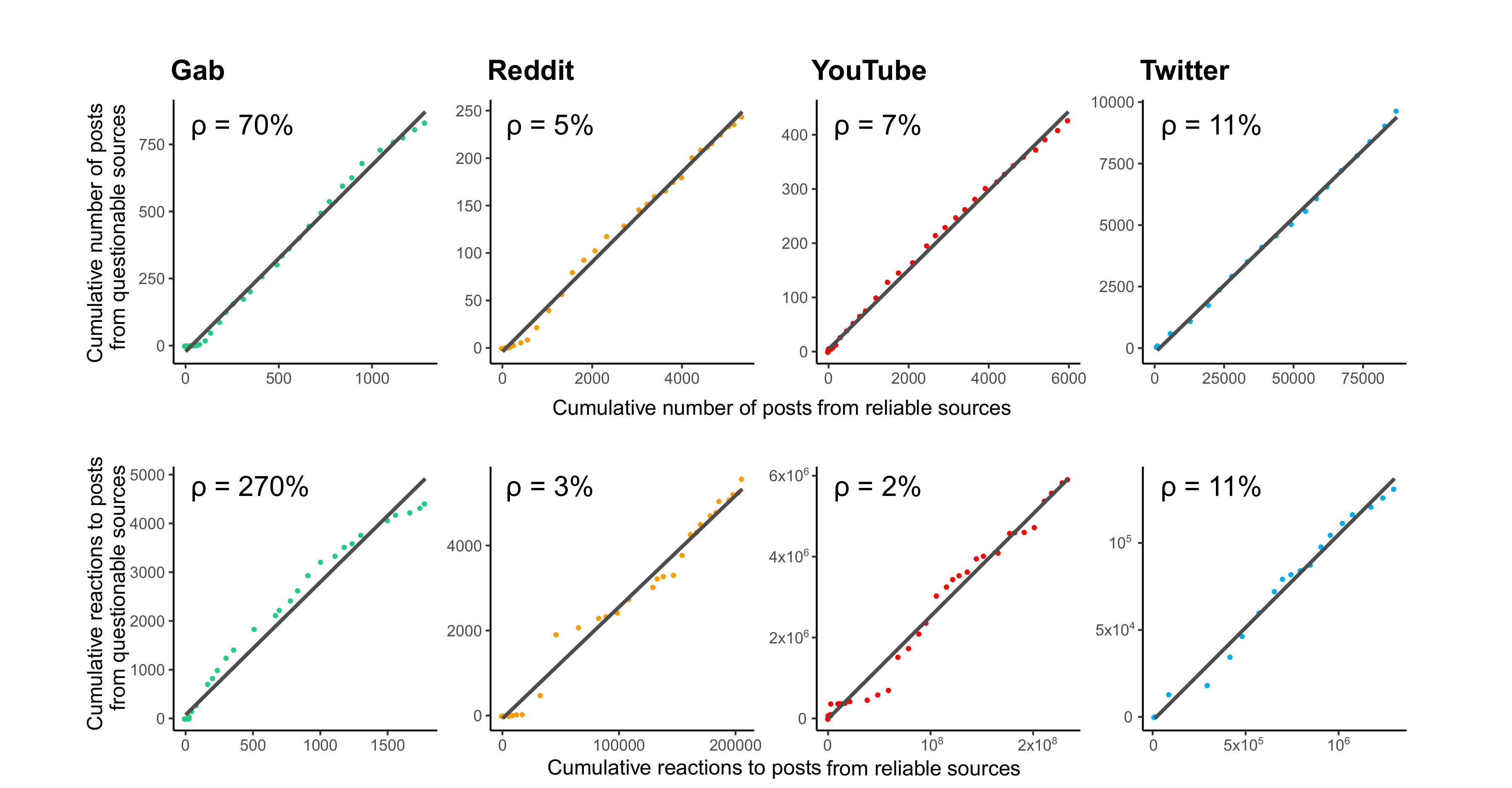}
    \caption{Upper panels: plot of the cumulative number of posts referring to questionable sources versus the cumulative number of posts referring to reliable sources. Lower panel: plot of the cumulative number of engagements relatives to questionable sources versus the cumulative number of engagements relative to reliable sources. Notice that a linear behavior indicates that the time evolution of questionable posts/engagements is just a re-scaled version of the time evolution of reliable posts/engagements. Each plot indicates the regression coefficients $\rho$, representing the ratio among the volumes of questionable and reliable posts ($\rho^{post}$) and engagements ($\rho^{eng}$). In more popular social media, the number of unreliable posts represent a small fraction of the reliable ones; same thing happens in Reddit. Among less popular social media, a peculiar effect is observed in Gab: while the volume of unreliable post is just the $\sim 70\%$ of the volume of reliable ones, the volume of engagements for unreliable posts is $\sim 270\%$ bigger than the volume for reliable ones. Further details concerning the regression coefficients are reported in SI.}
    \label{fig:3}
\end{figure}

Figure~\ref{fig:3} shows, for each platform, the plots of the cumulative number of posts and reactions related to questionable sources versus the cumulative number of posts and interactions referring to reliable sources. By interactions we mean the overall reactions, e.g. likes or other form or endorsement and comments, that can be performed with respect to a post on a social platform. Surprisingly, all the posts show a strong linear correlation, i.e., the number of posts/reactions relying on questionable and reliable sources grows with the same pace inside the same social media platform. We observe the same phenomenon also for the engagement with reliable and unreliable posts. Hence, the growth dynamics of unreliable posts/interactions is just a re-scaled version of the growth dynamics of reliable posts/reactions; however, the re-scaling factor $\rho$ (i.e., the fraction of unreliable over reliable) is strongly dependent on the platform. In particular, we observe that in mainstream social media the number of unreliable posts represent a small fraction of the reliable ones; the same thing happens in Reddit. Among less regulated social media, a peculiar effect is observed in Gab: while the volume of unreliable post is just the $\sim 70\%$ of the volume of reliable ones, the volume of reactions for unreliable posts is $\sim 270\%$ bigger than the volume for reliable ones. Such results hint the possibility that different platform react differently to reliable and unreliable news. 

To further investigate this issue, we define the amplification factor $\mathcal{E}$ as the average number of reactions to a post; hence, $\mathcal{E}$ is a measure that quantifies the extent to which a post is amplified in a social media. We observe that the amplification  $\mathcal{E}^U$ (for unreliable posts) and $\mathcal{E}^R$ (for reliable posts) varies from social media to social media and that assumes the largest values in YouTube and the lowest in Gab. To measure the permeability of a platform to reliable/unreliable news, we then define the coefficient of relative amplification $\alpha=\mathcal{E}^U/\mathcal{E}^R$. It is a measure of whether a social media amplifies questionable ($\alpha>1$) or reliable ($\alpha<1$) posts. Results are shown in Table \ref{tab:AmplificationRQ}. Among mainstream social media, we notice that Twitter is the most neutral ($\alpha \sim 100\%$ i.e.  $\mathcal{E}^U \sim \mathcal{E}^R$), while YouTube cuts out unreliable sources ($\alpha\sim 10\%$). Among less popular social media, Reddit reduces the impact of unreliable sources ($\alpha \sim 50\%$), while Gab strongly amplifies them ($\alpha\sim 400\%$).

Overall, our findings suggest that the main drivers of information spreading are related to specific peculiarities of each platform and depends upon the group dynamics of individuals engaged with the topic.

\begin{table}[H]
    \centering
    \begin{tabular}{|r|r|r|r|}\hline
    & $\mathcal{E}^U$ & $\mathcal{E}^R$ & $\alpha$ \\
\hline     
   Gab     & 5.6 & 1.4 & 3.9 \\
   Reddit  & 22.7 & 40.1 & 0.55 \\
   YouTube & 1.4$\times 10^4$ & 3.9$\times 10^4$ & 0.35 \\
   Twitter & 15.1 & 15.6 & 0.97\\
\hline
    \end{tabular}
    \caption{The average engagement of a post is the number reaction expected for a post and is a measure of how much a post is amplified in a social media. The average engagements  $\mathcal{E}^U$ (for unreliable post) and $\mathcal{E}^R$ (for reliable post) vary from social media to social media, and are the largest in Twitter and the lowest in Gab. The coefficient of relative amplification $\alpha=\mathcal{E}^U/\mathcal{E}^R$ measures whether a social media amplifies more unreliable ($\alpha>1$) or reliable ($\alpha<1$) posts. Among more popular social media, we notice that Twitter is the most neutral social media ($\alpha \sim 100\%$ i.e. $\mathcal{E}^U \sim \mathcal{E}^R$) while YouTube cuts out unreliable sources ($\alpha\sim 10\%$). Among less popular social media, Reddit reduces the impact of unreliable sources ($\alpha \sim 50\%$) while Gab strongly amplifies them ($\alpha\sim 400\%$).}
    \label{tab:AmplificationRQ}
\end{table}

\section{Conclusions}

In this work we perform a comparative analysis on five different social media platforms during the COVID-19 health emergency.
Such a timeframe is good benchmark for studying content consumption dynamics around critical events in a historic times when the accuracy of information is threatened. 
We assess users engagement and interest about the COVID-19 topic and characterize the evolution of the discourse over time. Furthermore, we model the spread of information with epidemic models by providing basic growth parameters for each social media. 
We then analyze the spreading of questionable information for all channels, finding that Gab is the environment more susceptible to misinformation diffusion. However, information marked either as reliable or questionable do not present significant differences in their spreading patterns.
Our analysis suggests that information spreading is driven by the interaction paradigm imposed by the specific social media or/and by the specific interaction patterns of groups of users engaged with the topic. Finally, we conclude the paper by providing COVID-19 rumor’s amplification parameters for social media platform.
%We conclude the paper by providing amplification parameters for each social media.
We believe that the understanding of social dynamics behind content consumption and social media is an important subject, since it may help to design more efficient epidemic models accounting for social behavior and to implement more efficient communication strategies in time of crisis.

\section{Materials and Methods}
\subsection{Data Collection}

Table \ref{tab:databreak} reports the data breakdown of the five social platforms. Given the diversity of social media platforms, five different data collection processes have been performed. For Gab, Reddit, YouTube and Twitter data were gathered by the existing API services, while for Instagram no API services were available thus we built our own process. In particular, we manually collected data by visual inspection to build up the database for the analysis.

%
%Reddit
Reddit dataset was downloaded from the Pushift.io\footnote{\url{https://pushshift.io/}} archive, exploiting the related API. In order to filter contents linketo COVID-19, we selected a group of keywords based on Google Trends' COVID-19 related queries such as: coronavirus, pandemic, coronaoutbreak, china, wuhan, nCoV, IamNotAVirus, coronavirus\_update, coronavirus\_transmission, coronavirusnews, coronavirusoutbreak.

%
%Gab
In Gab, although no official guides are available, there is an API service that given a certain keyword, returns a list of users, hashtags and groups related to it. We queried all the keywords we selected based on Google Trends and we downloaded all hashtags linked to them. We then manually browsed the results and selected a set of hashtags based on their meaning. For each hashtag in our list, we downloaded all the posts and comments linked to it.

%
%YouTube
For YouTube, we collected videos by using the YouTube Data API\footnote{\href{https://developers.google.com/youtube/v3/docs}{Link: YouTube Data API}} by searching for videos that matched a variety of queries: coronavirus $OR$ coronavirus weapon $OR$ coronavirus epidemic $OR$ coronavirus outbreak $OR$ coronavirus pandemic $OR$ coronavirus conspiracy $OR$ coronavirus news $OR$ nCov-2019 $OR$ \#coronavirus $OR$ nCov $OR$ corona virus $OR$ corona-virus $OR$ novelcoronavirus $OR$ wuhanvirus $OR$ novel coronavirus $OR$ wuhan virus $OR$ coronavirus bio-weapon $OR$ corvid-19 $OR$ COVID-19. Then an in depth search was done by crawling the network of videos by searching for more related videos as established by the YouTube algorithm. From the gathered set, we filtered the videos that contained coronavirus $OR$ nCov $OR$ corona virus $OR$ corona-virus $OR$ corvid $OR$ covid $OR$ SARS-CoV in the title or description. We also limited the dataset to the videos published during the analysis period (January 1 to February 14). We then collected all the comments received by those videos. This was done on February 28.
%
%Twitter
For Twitter, we collect tweets related to the topic corona-virus by using both Twitter the search and stream endpoint of the Twitter API\footnote{\href{https://developer.twitter.com/en.html}{Link: Twitter API}} using the following queries: coronavirus $OR$ ncov $OR$ coronavirusoutbreak $OR$ wuhan $OR$ iamnotavirus. The data deriving from stream API represent only 1\% of the total volume of tweets, further filtered by the selected keywords.The data derived from the search API represent a random sample of the tweets containing the selected keywords up to a maximum rate limit of 18000 tweets every 10 minutes. 
%
%Instagram
Since no official API are available for Instagram data, we built our own process to collect public contents related to specific keywords such as: coronavirus, nCov, wuhan, pandemic and imnotavirus. We manually took notes of posts, comments and populated the Instagram Dataset.

The results related to the engagement of users are obtained using only API search results. 

\begin{table}[h]
     \centering
     \begin{tabular}{|c|c|c|c|c|} \hline
          & \textbf{Posts} & \textbf{Comments} & \textbf{Users} & \textbf{Period} \\
          \hline     
\textbf{Gab} & 6,252 & 4,364 & 2,629 & 01/01-14/02 \\   
\textbf{Reddit} &  10,084 & 300,751 & 89,456 & 01/01-14/02 \\
\textbf{YouTube} & 111,709 & 7,051,595 & 3,199,525 & 01/01-14/02 \\
\textbf{Instagram} & 26,576 & 109,011 & 52,339 & 01/01-14/02 \\
\textbf{Twitter} & 1,187,482 & - & 390,866 & 27/01-14/02 \\
 \hline 
\textbf{Total} &  1,342,103 & 7,465,721 & 3,734,815 &  \\
\hline
     \end{tabular}
     \caption{Data breakdown of the number of posts, comments and users for all platforms.}
     \label{tab:databreak}
 \end{table}

 \subsection{Text analysis}\label{subsec:Clustering_procedure}
 
To provide an overview of the debate concerning the virus outbreak on the various platforms, we extract and analyze all topics related to COVID-19 by applying Natural Language Processing techniques to the written content of each social media. We first build word embedding for the text corpus of each platform, then, to assess the topics around which the perception of the COVID-19 debate is concentrated, we cluster words by running the Partitioning Around Medoids (PAM) algorithm on their vector representations.
 
Word embeddings, i.e., distributed representations of words learned by neural networks, represent words as vectors in ${\mathbf R}^n$ bringing similar words closer to each other. They perform significantly better than the well-known Latent Semantic Analysis (LSA) and Latent Dirichlet Allocation (LDA) for preserving linear regularities among words and computational efficiency on large data sets \cite{Mikolov2013b}. In this paper we use the Skip-gram model \cite{Mikolov2013} to construct word embedding of each social media corpus. More formally, given a content represented by the sequence of words $w_1,w_2,\dots,w_T$, we use stochastic gradient descent with gradient computed through backpropagation rule \cite{Rumelhart1986} for maximizing the average log probability
\begin{equation}
\frac{1}{T}\displaystyle\sum_{t=1}^T\left[\displaystyle\sum_{j=-k}^k\log p(w_{t+j}\vert w_t)\right]
\end{equation}
where $k$ is the size of the training window. Therefore, during training the vector representations of closely related words are pushed to be close to each other.

In the Skip-gram model, every word $w$ is associated with its input and output vectors, $u_w$ and $v_w$, respectively. The probability of correctly predicting the word $w_i$ given the word $w_j$ is defined as
\begin{equation}
p(w_i\vert w_j)=\frac{\exp\left(u_{w_i}^Tv_{w_j}\right)}{\displaystyle\sum_{l=1}^V\exp\left(u_{l}^Tv_{w_j}\right)}
\end{equation}
where $V$ is the number of words in the corpus vocabulary. Two major parameters affect the training quality: the dimensionality of word vectors, and the size of the surrounding words window. We choose 200 as vector dimension -- that is typical value for training large dataset -- and 6 words for the window.

Before applying the tool, we reduced to contents written in English language as detected with \texttt{cld3}\footnote{ https://cran.r-project.org/web/packages/cld3/index.html}. Then we cleaned the corpora by removing HTML code, URLs and email addresses, user mentions, hashtags, stop-words, and all the special characters including digits. Finally, we dropped words composed by less than three characters, words occurring less than five times in all the corpus, and contents with less than three words.

To analyze the topics related to COVID-19, we cluster words by Partitioning Around Medoids (PAM) and using as proximity metric the cosine distance matrix of words in their vector representations. In order to select the number of clusters, $k$, we calculate the average silhouette width for each value of $k$. Moreover, for evaluating the cluster stability, we calculate the average pairwise Jaccard similarity between clusters based on 90\% sub-samples of the data. Lastly, we produce word clouds to identify the topic of each cluster. 
To provide a view about the debate around the virus outbreak, we define the distribution over topics $\Theta_c$ for a given content $c$ as the distribution of its words among the word clusters. Thus, to quantify the relevance of each topic within a corpus, we restrict to contents $c$ with $\max\Theta_c>0.5$ and consider them uniquely identified as a single topic each.

\begin{table}[h]
     \centering
     \begin{tabular}{|c|c|c|c|c|} \hline
          & \textbf{Cleaned contents} & \textbf{Vocabulary size} & \textbf{Topics} & \textbf{Contents with $\max\Theta>0.5$} \\
          \hline        
\textbf{Instagram} & 21,189 posts & 15,324 & 17 & 4,467 \\
\textbf{Twitter} & 638,214 posts & 22,587 & 21 & 369,131 \\
\textbf{Gab} & 5,853 posts & 3,024 & 19 & 2,986 \\
\textbf{Reddit} &  10,084 posts & 1,968 & 34 & 6,686 \\
\textbf{YouTube} & 815,563 comments & 35,381 & 30 & 679,261\\
\hline
     \end{tabular}
     \caption{Results of text cleaning and analysis for all the corpora.}
     \label{tab:cleaneddata}
 \end{table}

\subsection{Epidemiological Models}\label{subsec:MM-epidemics}

To allow predictions,  assess the impact of policies, and thus to define optimal control/communication strategies, it is important to model the dynamics observed from data. Several mathematical models can be used to analyse potential mechanisms that underline epidemiological data; generally, we can distinguish among phenomenological models that emphasize the reproducibility of empirical data without insights in the mechanisms of growth, and more insightful mechanistic models that try to incorporate such mechanisms \cite{Chowell2017}.  

To fit our cumulative curves, we first use the adjusted exponential model of \cite{Fisman2013} since it naturally provides an estimate of the basic reproduction number $R_0$. This phenomenological model (from now on indicated as EXP) has been successfully employed in data-scarce settings and shown to be on-par with more traditional compartmental models for multiple emerging diseases like Zika, Ebola, and Middle East Respiratory Syndrome \cite{Fisman2013}.

The model is defined by the following single equation:

\begin{equation}
I = \left[ \frac{R_0}{(1+d)^t} \right]^t
\label{eq:EXP}    
\end{equation}
Here, $I$ is incidence, $t$ is the number of days, $R_0$ is the basic reproduction number and $d$ is a damping factor accounting for the reduction in transmissibility over time.% due to the natural depletion of susceptible individuals in the affected population and any public interventions that may impact disease spread over time. 
In our case, we interpret $I$ as the number $C_{auth}$ of authors that have published a post on the subject.

As a mechanistic model, we employ the classical SIR model \cite{Bailey1975book}. In such a model, a susceptible population can be infected with a rate $\beta$ by coming into contact with infected individuals; however, infected individuals can recover with a rate $\gamma$. The model is described by a set of differential equations:
%\begin{equation}
\begin{align}
\notag \partial_t S &= - \beta S \cdot I /N \\
\partial_t I& =  \beta S \cdot I /N - \gamma I \\
\notag \partial_t R &= \gamma  I 
\end{align}
\label{eq:SIR}    
%\end{equation}
where $S$ is the number of susceptible, $I$ is the number of infected and $R$ is the number of recovered. In our case, we interpret the number $I+R$ as the number $C_{auth}$ of authors that have published a post on the subject.

In the SIR model, the basic reproduction number $R_0=\beta/\gamma$ corresponds to the ration among the rate of infection by contact $\beta$ and the rate of recovery $\gamma$. Notice that for the SIR model, vaccination strategies correspond to bringing the system in a situation where $S<N/R_0$; in such a way, both the number of infected will decrease. 

To estimate the basic reproduction numbers $R_0^{EXP}$ and $R_0^{SIR}$ for the EXP and the SIR model, we use least square estimates of the models' parameters\cite{Ma2020}. The range of parameters is estimated via bootstrapping \cite{Efron1994book,Chowell2017}.

\subsection{Regression Table of Figure 3}

Table~\ref{tab:RQfits} reports the regression coefficients and $R^2$ values displayed in Figure~\ref{fig:3}. We observe an overall high value of $R^2$ meaning a strong explanatory power of the performed linear regressions.

 \begin{table}[h]
     \centering
     \begin{tabular}{|c|c|c|c|c|} \hline
          & Name & Intercept & Coefficient ($\rho$)& $R^2$ \\
          \hline
          
Regression 1 & Gab users & -6.163 & 0.80  & 0.992 \\
Regression 2 & Reddit users & -6.439 & 0.085  & 0.995 \\
Regression 3 & YouTube users & 0.890 & 0.149  & 0.999 \\
Regression 4 & Twitter users & -118.157 & 0.121 & 0.991 \\
Regression 5 & Gab posts & -22.321 & 0.695 & 0.996 \\
Regression 6 & Reddit posts & -4.111 & 0.047 & 0.997 \\
Regression 7 & YouTube posts & 4.529 & 0.073 & 0.998 \\
Regression 8 & Twitter posts & -151.44 & 0.110 & 0.998 \\
Regression 9 & Gab interactions & 74.577 & 2.721 & 0.981 \\
Regression 10 & Reddit interactions & -70.677 & 0.026 & 0.990 \\ 
Regression 11 & YouTube interactions & -8854.33 & 0.025 & 0.986 \\
Regression 12 & Twitter interactions & -2136.978 & 0.107 & 0.987 \\
\hline
     \end{tabular}
     \caption{Coefficients and $R^2$ of the linear regressions displayed in Figure 3.}
     \label{tab:RQfits}
 \end{table}

\subsection{Matching ability}

We consider all the posts in our dataset that contain at least one Uniform Resource Locator (URL) linking to a website outside the related social media (e.g., tweets pointing outside Twitter). We separate URLs in two main categories obtained using the classification provided by MediaBias/FactCheck (MBFC). MBFC provides a classification determined by ranking bias in four different categories that are Biased Wording/Headlines, Factual/Sourcing, Story Choices and Political Affiliation. A score is assigned to each category per each news outlet and the average score determined the bias of the outlet, as explained in the Methodology Section of the website.

Accordingly, to each news outlet is associated a label that refers either to a political bias, namely, Right, Right-Center, Least-Biased, Left-Center and Left or to its reliability that is expressed in three labels namely, Conspiracy-Pseudoscience, Pro-Science or Questionable. Noticeably, also the Questionable set include a wide range of political bias, from Extreme Left to Extreme Right. For instance, the Right label is associated to Fox News, the Questionable label to Breitbart (the well-known extreme right outlet) and the Pro-Science label to Science. Using such a classification, we assign to each of these outlets a binary label that partially stems from the labelling provided by MBFC. Indeed, we divide the news outlets into Questionable outlets and Reliable outlets. All the outlets already classified as Questionable or belonging to the category Conspiracy-Pseudoscience are labelled as Questionable, the rest is labelled as Reliable.   

Considering all the 2637 news outlets that we retrieve from the list provided by MBFC we end up with 800 outlets classified as Questionable 1837 outlets classified as Reliable.

Using such a classification we quantify our overall ability to match and labels domains of posts containing URLs (that are expanded in case they come in their shortened form). Our matching ability is reported in Table~\ref{tab:match}.

\begin{table}[H]
    \centering
    \begin{tabular}{|c|c|c|c|c|c|}\hline
        & Gab & Reddit & YouTube & Instagram & Twitter  \\ \hline
Posts containing a URL  & 3778 & 10084 & 351786 & 1328 & 356448 \\
Matched & 0.47 & 0.55 & 0.035 & 0.09 & 0.27 \\ 
Questionable & 0.38 & 0.045 & 0.064 & 0.05 & 0.10 \\
Reliable & 0.62 & 0.955 & 0.936 & 0.95 & 0.90 \\ \hline
    \end{tabular}
    \caption{Number of posts containing a URL, matching ability and classification for each of the five platforms.}
    \label{tab:match}
\end{table}

The matching ability that, in some cases like YouTube is pretty low, doesn't refer to the ability of identifying known domain but to the ability of finding the news outlets that belong to the list provided by MBFC. Indeed in all the social networks we find a strong tendency towards linking to other social media post that we are unable to match. The percentage of inter and intra-linking of social media is reported in Table \ref{tab:intralinks}.

\begin{table}[H]
    \centering
    \begin{tabular}{|c|c|c|c|c|c|c|}\hline
& Gab & Reddit & YouTube & Instagram & Twitter & Facebook \\ \hline
Gab & 0.003 & 0.002 & 0.001 & 0.002 & 0.138 & $\sim$0 \\
Reddit & 0.043 & 0.006 & 0.009& 0.001 & $\sim$0 & 0 \\
YouTube & 0 & $\sim$0 & 0.292 & $\sim$0 & 0.088 & 0.081 \\
Instagram & 0 & 0 & 0.003 & 0 & 0.001 & 0.001 \\
Twitter & 0.059 & 0.001 & 0.257 & 0.003 & $\sim$0 & $\sim$0 \\ \hline
\end{tabular}
\caption{Fraction of URLs pointing to social media.}
\label{tab:intralinks}
\end{table}

\bibliographystyle{unsrt}
\bibliography{covid19}

\end{document}